# Vaccination, life expectancy, and trust: Patterns of COVID-19 and measles vaccination rates around the world

Cosima Rughiniș, Department of Sociology, University of Bucharest, Bucharest, 030167, Romania, cosima.rughinis@unibuc.ro

Simona - Nicoleta Vulpe, Interdisciplinary School of Doctoral Studies, University of Bucharest, Bucharest, 050107, Romania, simona.vulpe@drd.unibuc.ro

Michael G. Flaherty, Department of Sociology, Eckerd College, FL 33711 USA, flahermg@eckerd.edu

Sorina Vasile, Interdisciplinary School of Doctoral Studies, University of Bucharest, Bucharest, 050107, Romania, sorina.vasile@gmail.com

**Corresponding author:**

Simona - Nicoleta Vulpe, 36-46 Mihail Kogălniceanu, room 221 first floor, Sector 5, Bucharest, 050107, Romania. E-mail: simona.vulpe@drd.unibuc.ro.

# Vaccination, life expectancy, and trust: Patterns of COVID-19 and measles vaccination rates around the world

## Abstract

We estimate patterns of covariation between COVID-19 and measles vaccination rates and a set of widely used indicators of human, social, and economic capital across 146 countries. About 70% of the variability in COVID-19 vaccination rates in February 2022, worldwide, can be explained by differences in the Human Development Index (HDI) and, specifically, in life expectancy at birth. Trust in doctors and nurses adds predictive value beyond the HDI, clarifying controversial discrepancies between vaccination rates in countries with similar levels of human development and vaccine availability. Cardiovascular disease deaths, an indicator of general health system effectiveness, and infant measles immunization coverage, an indicator of country-level immunization effectiveness, are also significant, though weaker, predictors of COVID-19 vaccination success. The metrics of economic inequality, perceived corruption, poverty, and inputs into the health system have strong bivariate correlations with COVID-19 vaccination but no longer remain statistically significant when controlling for the HDI. Measles vaccination in 2019 is similarly predicted by HDI, trust in doctors and nurses. National poverty rates seem to be a relevant predictor for both types of vaccination, though statistical significance is oscillating. The remaining variability in COVID-19 vaccination success that cannot be pinned down through these sets of metrics points to a considerable scope for collective and individual agency in a time of crisis. The mobilization and coordination in the vaccination campaigns of citizens, medical professionals, scientists, journalists, and politicians, among others, account for at least some of this variability in overcoming vaccine hesitancy and inequity.

## Introduction

Comedian Dave Barry recalled his mother telling him, “Son, it is better to be rich and healthy than poor and sick” [1]. This still holds when examining COVID-19 vaccination patterns worldwide. In this paper, we discuss the relative contribution to predicting COVID-19 and measles vaccination rates of a set of widely used, publicly available indicators of human, social, and economic capital.

There has been a significant increase in life expectancy over the last two hundred years in many societies. Humankind has become more adept, collectively, to sustain life for its members, although externalities, in terms of climate impact, have begun to raise doubt on the longer-term perspectives of this accomplishment. Life expectancy serves as a synthetic measure of the capacity of society to prevent death in a certain period. Given that the avoidance of death is one of humankind’s major goals, life expectancy is, therefore, a useful metric to capture the effectiveness of social organization for public health at a certain time and place.

*Vaccination has played a considerable role in reducing the mortality inflicted by preventable diseases* [2] over the last two centuries. Vaccines have been, therefore, an important cause of the recent increase in life expectancy across the world. This also holds true for the COVID-19 pandemic, which has visibly lowered life expectancy in most countries [3], [4]. There is convincing evidence that vaccination against COVID-19 has prevented numerous deaths globally [5].

At the same time, *rates of vaccination have varied widely during the pandemic*. *Societal resources shape a collectivity’s ability to immunize its members against infection through vaccination* [6]*.* However, COVID-19 vaccination has been unevenly implemented because of differences in availability of vaccines, uneven logistics of vaccine distribution, and people’s variable trust in vaccines and mainstream science and expertise [7], [8], [9], [10], [11]. In this paper, we explore and discuss the correlation between the success of vaccination campaigns against COVID-19 in mid-2021 and early 2022 and pre-pandemic life expectancy (estimated in 2019), alongside other measures of human, social, and economic capital, at country-level. Our study aimed to answer an essential question: What can such broad patterns of co-variation in vaccination success tell us about the social structures and forms of agency that keep people alive?

Human, social, and economic resources have been of utmost importance in COVID-19 vaccination. They have facilitated earlier access to vaccines and powered the required logistics of a large-scale vaccination campaign. Several studies signaled a positive association between coverage of COVID-19 vaccination, the Human Development Index (HDI), and gross domestic product (GDP) per capita [12], [13], [14]. Education and GDP per capita also contribute to the speed of the COVID-19 vaccination campaign [15]. A positive correlation between measles vaccination and HDI has also been noted [16]. Trust in the state and in the health system has been associated with greater compliance with COVID-19 restrictions in Europe [17]. Trust in medical and scientific experts has been a strong correlate of pro-vaccination attitudes in general [18], [19], [20], [21] and of the declared intention to receive a COVID-19 vaccine internationally [22], [23], [24]. Social and economic inequality has been associated with lower COVID-19 vaccination rates aggregated at country-level [25]. Indicators of corruption in the public sector are significant predictors of COVID-19 vaccination in August 2021 when controlling for GDP per capita and strength of the health system [26], without controlling for life expectancy or education. Perceived corruption is associated with decreased vaccination coverage globally [27] and it also affects trust in mainstream health policy, exacerbating vaccination hesitancy [28].

## Methods

We accessed publicly available data on COVID vaccination rates and other country-level indicators of human, social, and economic capital from the datasets of Our World in Data (OWID) [29], the metrics included in the 2020 Human Development Report (HDR) of the United Nations Development Programme [30], the Corruption Perception Index computed by Transparency International [31], and World Bank data on poverty rates [32]. We included in the study all countries and territories with a population larger than 1 million and available information for vaccination rates, according to OWID data, resulting in 146 units of analysis[1].

Our first dependent variable of interest was the rate of fully vaccinated people, per hundred, measured at two points in time: July 31, 2021 (or the closest day to July 31, 2021) and February 4, 2022 (or the closest day to February 4, 2022). The second dependent variable, included for comparison purposes, is the rate of infants immunized against measles at one year of age, in 2019, as reported by the HDR. The descriptive statistics and sources for the predictors included in the analysis are presented in Table 1. The control variable for partial correlations was the HDI, which aggregates three dimensions: 1) life expectancy at birth; 2) an education index composed of mean years of schooling and expected years of schooling; and 3) gross national income per capita (GNI) [30].

*Table 1. Descriptive statistics for the variables included in the analysis. Source: Authors' analysis on publicly available data from OWID, UNDP HDR, Transparency International, and The World Bank*

| | Number of cases (N) | Minimum | Maximum | Mean | Std. Deviation |
|---|---|---|---|---|---|
| People fully vaccinated per hundred, Feb. 2022 (OWID) | 146 | 0.23 | 93.55 | 44.50 | 27.59 |
| People fully vaccinated per hundred, July 2021 (OWID) | 144 | 0.01 | 66.00 | 15.65 | 17.83 |
| Human Development Index 2019 (UNDP HDR) | 152 | 0.36 | 0.96 | 0.72 | 0.16 |
| Life expectancy at birth 2019 (HDI component, UNDP HDR) | 153 | 53.28 | 84.86 | 72.52 | 7.85 |
| Mean years of schooling 2019 (HDI component, UNDP HDR) | 151 | 1.64 | 14.15 | 8.74 | 3.19 |

[1] The countries included in the analysis are, in alphabetical order: Afghanistan, Albania, Algeria, Angola, Argentina, Armenia, Australia, Austria, Azerbaijan, Bahrain, Bangladesh, Belarus, Belgium, Benin, Bolivia, Bosnia and Herzegovina, Botswana, Brazil, Bulgaria, Burkina Faso, Cambodia, Cameroon, Canada, Central African Republic, Chad, Chile, China, Colombia, Costa Rica, Cote d'Ivoire, Croatia, Cuba, Czech Republic, Democratic Republic of Congo, Denmark, Dominican Republic, Ecuador, Egypt, El Salvador, Equatorial Guinea, Estonia, Eswatini, Ethiopia, Finland, France, Gabon, Gambia, Georgia, Germany, Ghana, Greece, Guatemala, Guinea, Guinea-Bissau, Honduras, Hong Kong, Hungary, India, Iran, Iraq, Ireland, Israel, Italy, Jamaica, Japan, Jordan, Kazakhstan, Kenya, Kuwait, Kyrgyzstan, Laos, Latvia, Lebanon, Liberia, Libya, Lithuania, Madagascar, Malawi, Malaysia, Mali, Mauritania, Mexico, Moldova, Mongolia, Morocco, Mozambique, Myanmar, Namibia, Nepal, Netherlands, New Zealand, Nicaragua, Niger, Nigeria, North Macedonia, Norway, Oman, Pakistan, Palestine, Panama, Papua New Guinea, Paraguay, Peru, Philippines, Poland, Portugal, Romania, Russia, Rwanda, Saudi Arabia, Senegal, Serbia, Sierra Leone, Singapore, Slovakia, Slovenia, Somalia, South Africa, South Korea, South Sudan, Spain, Sri Lanka, Sudan, Sweden, Switzerland, Syria, Taiwan, Tajikistan, Thailand, Timor, Togo, Trinidad and Tobago, Tunisia, Turkey, Uganda, Ukraine, United Arab Emirates, United Kingdom, United States, Uruguay, Uzbekistan, Venezuela, Vietnam, Yemen, Zambia, Zimbabwe.

| | | | | | |
|---|---|---|---|---|---|
| Expected years of schooling 2019 (HDI component, UNDP HDR) | 148 | 5.30 | 21.95 | 13.43 | 3.01 |
| GNI per capita 2019 in 2017 PPP (HDI component, UNDP HDR) | 148 | 993.01 | 92,418.23 | 20,129.66 | 20,023.66 |
| Atkinson Index of inequality in life expectancy (UNDP HDR) | 149 | 2.50 | 40.90 | 14.72 | 10.68 |
| Atkinson Index of inequality in education (UNDP HDR) | 143 | 0.70 | 50.12 | 18.85 | 14.45 |
| Atkinson Index of inequality in income (UNDP HDR) | 131 | 8.50 | 57.00 | 23.03 | 9.28 |
| Gini Index on inequality in income 2019 (UNDP HDR) | 138 | 0.00 | 63.03 | 36.82 | 9.69 |
| PISA Programme for International Student Assessment Score Reading 2018 (OWID, from OECD) | 67 | 339.69 | 555.24 | 453.32 | 53.78 |
| PISA Programme for International Student Assessment Score Mathematics 2018 (OWID, from OECD) | 68 | 325.10 | 591.39 | 456.79 | 56.78 |
| PISA Programme for International Student Assessment Score Science 2018 (OWID, from OECD) | 68 | 335.63 | 590.45 | 457.39 | 52.31 |
| World Bank - Poverty ratio (%) | 51 | 2.37 | 74.20 | 28.11 | 15.35 |
| World Bank - National poverty ratio (%) | 128 | 0.60 | 76.80 | 27.86 | 17.00 |
| Extreme poverty rate (%) (OWID) | 103 | 0.10 | 77.60 | 13.12 | 20.10 |
| Corruption Perception Index CPI 2020 (Transparency International) | 151 | 12.00 | 88.00 | 42.47 | 19.05 |
| Share of people who trust their national government 2018 (%) (OWID, from Wellcome Trust) | 124 | 10.95 | 99.22 | 51.51 | 18.11 |
| Share of people who trust doctors and nurses in their country 2018 (%) (OWID, from Wellcome Trust) | 134 | 43.43 | 98.20 | 80.79 | 11.56 |
| Health expenditure % of GDP in 2017 (UNDP HDR) | 144 | 1.18 | 17.06 | 6.63 | 2.61 |
| Physicians per 1000 people 2019 (UNDP HDR) | 145 | 0.23 | 84.22 | 19.31 | 16.94 |
| Hospital beds per 1000 people 2019 (UNDP HDR) | 135 | 1.00 | 129.80 | 28.95 | 24.33 |
| Cardiovascular Disease Death Rate per 100,000 people (OWID) | 145 | 79.37 | 724.42 | 260.84 | 123.43 |
| Diabetes Prevalence (%) (OWID) | 144 | 0.99 | 17.72 | 7.18 | 3.52 |
| Infants immunized for measles at 12 months, 2019 (%) (UNDP HDR) | 148 | 37.00 | 99.00 | 86.88 | 14.15 |

# Results

An exploration of bivariate correlations indicated a strong relationship between COVID-19 vaccination rates and the HDI (bivariate *r* = 0.826 in February 2022, *p* = 0.000). The relationship changed from an exponential to a linear shape during the vaccination campaign from July 2021 (Fig. 1) to February 2022 (Fig. 2). In mid-2021, there was a much more abrupt co-variation of vaccination success with HDI, compared with the later stage, when access to vaccines was more widespread and countries' own resources for large-scale collective action became more relevant.

Therefore, an exponential regression model ($R^2$ = 66.7%) is better fitted for the observed data in July than a linear regression model ($R^2$ = 48.3%). For February 2022, a linear model is better suited to model the relationship between HDI and vaccination rate ($R^2$ = 68.0%) than an exponential model ($R^2$ = 62.5%). A logarithmic model is marginally less fitted ($R^2$= 66%) than a linear one, anticipating a turn toward a logarithmic-shaped relationship as more countries on the HDI continuum evolve toward the plateau of high vaccination rates.

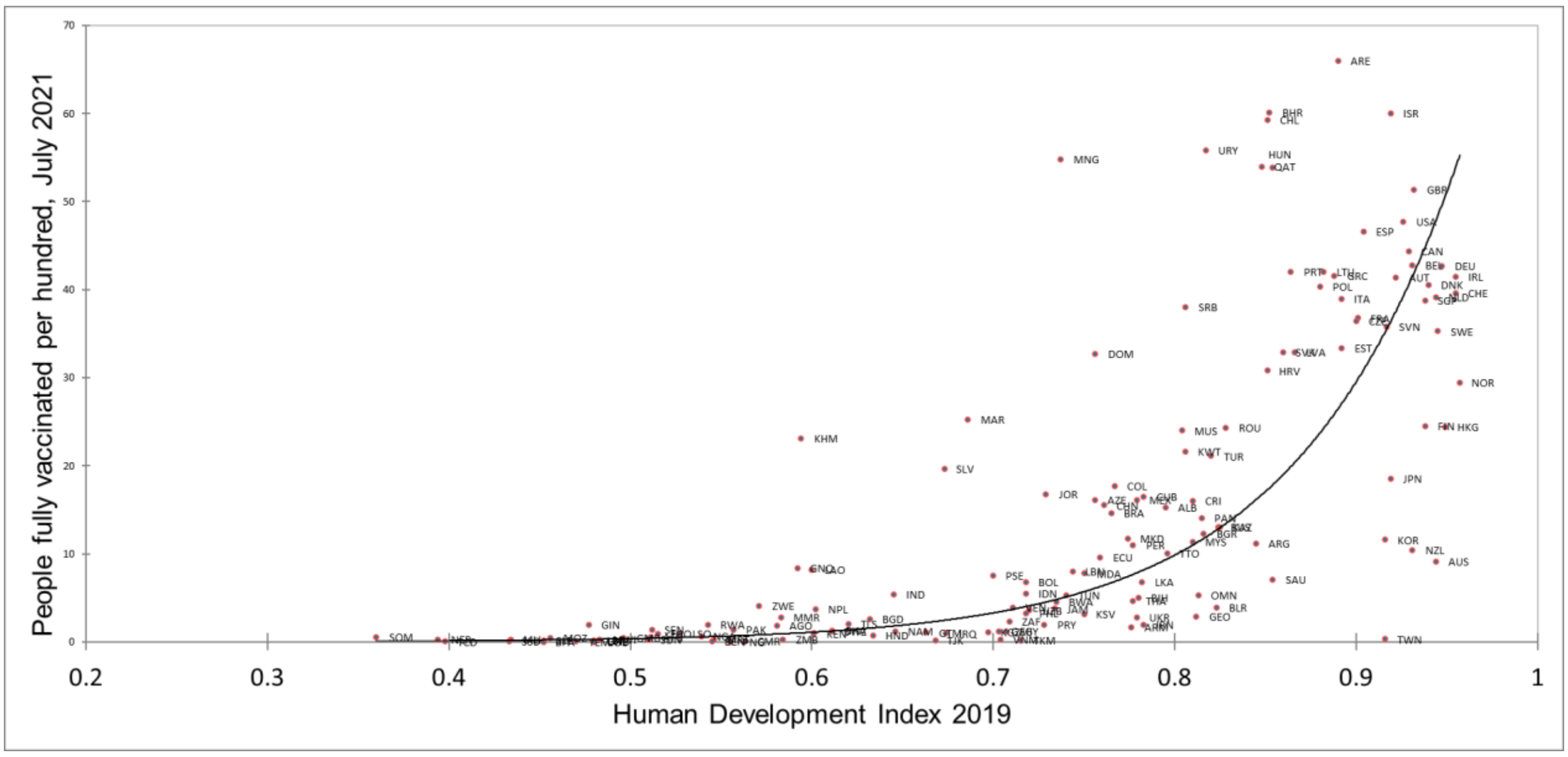

**Fig. 1.** Scatterplot of rates of fully vaccinated people in July 2021 vs. HDI 2019. Source: Authors' analysis of data from Our World in Data and UNDP Human Development Reports. Linear Pearson correlation: r = 0.695 (p = 0.000).

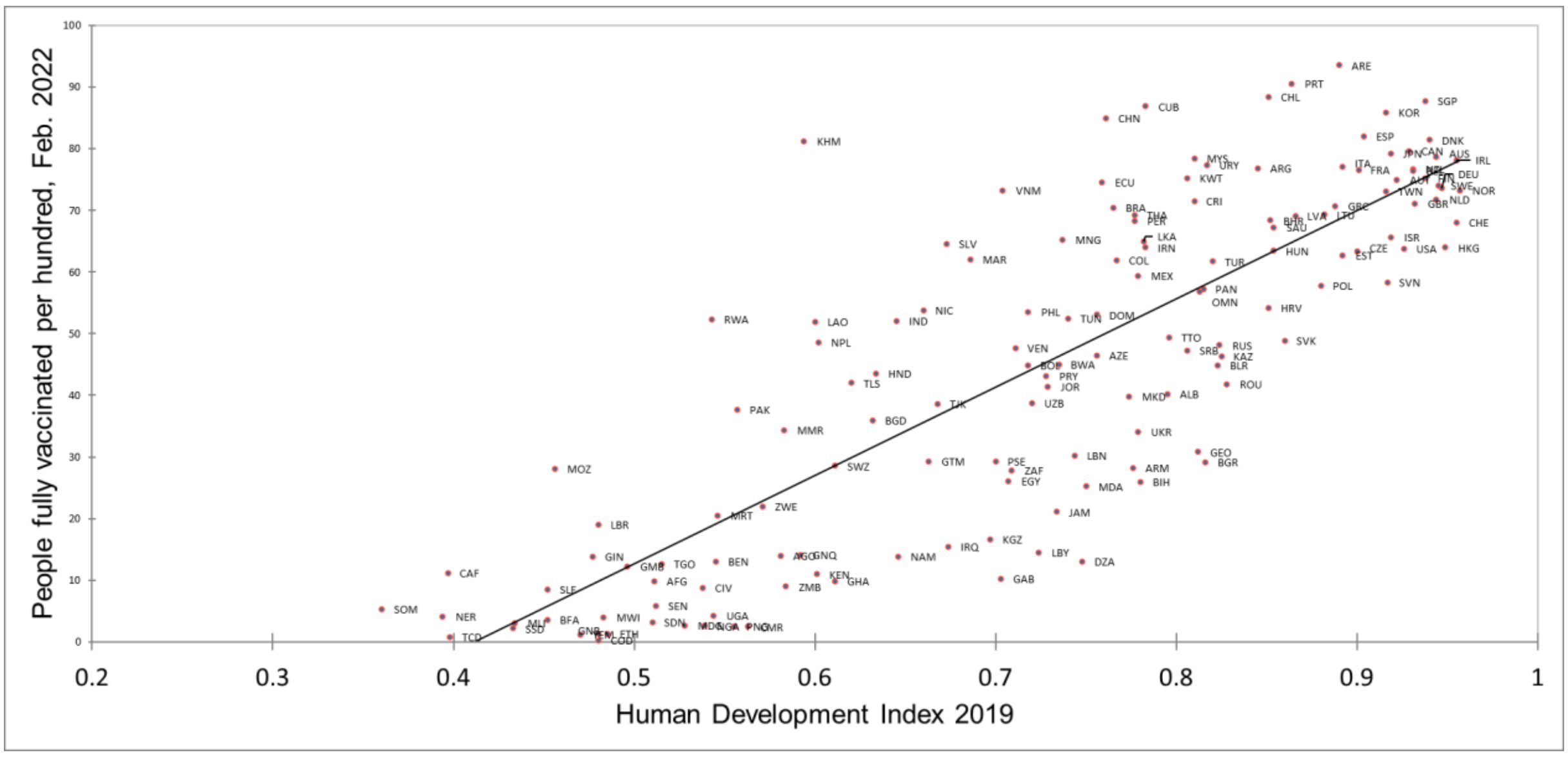

**Fig. 2.** Scatterplot of rates of fully vaccinated people in February 2022 vs. HDI 2019. Source: Authors' analysis of data from Our World in Data and UNDP Human Development Reports. Linear Pearson correlation: r = 0.826 (p = 0.000).

A bivariate analysis of vaccination rates and multiple indicators of human, social, and economic capital indicates a broad pattern of covariation (Table 2). *Vaccination rates are higher, on average, in countries with better outcomes in health and education, higher inputs into the health system, lower inequality, lower poverty rates, lower perceived corruption, and higher trust rates*.

The indicators that stand out in this pattern through their relative predictive power (other than the aggregate HDI) are *life expectancy at birth and GNI per capita*. Life expectancy at birth correlates at 0.836 with the vaccination rate in February 2022, explaining about 70% of its total variance.[2] GNI per capita correlates at 0.706 with vaccination rates in February 2022, explaining about 50% of the total variance, which makes it the second strongest predictor in the bivariate analysis. Mean years of schooling also correlates at 0.688 with the February 2022 vaccination rate.

*Table 2. Bivariate Bravais-Pearson correlations and partial correlations controlling for HDI 2019 between vaccination rates and indicators of human, economic, and social capital indicators. Source: Authors' analysis on publicly available data from OWID, UNDP HDR, Transparency International, and The World Bank*

| | **Bivariate correlations** | | | | | **Partial correlations when controlling for the HDI 2019** | | |
|---|---|---|---|---|---|---|---|---|
| Variables | Metric | People fully vaccinated per hundred, | People fully vaccinated per hundred, | Infants immunized for measles | Human Developme nt Index | Metric | People fully vaccina | Infants immuniz ed for |

[2] The inequality of life expectancy, estimated in the HDRs, is also strongly correlated with the vaccination rate, but it is collinear with the measure of life expectancy, and thus, it does not add predictive information.

| | | Feb. 2022 (OWID) | July 2021 (OWID) | for one-year olds, 2019 (HDR) | (HDI) 2019 | | ted per hundre d, Feb. 2022 (OWID | measles at 12 months, 2019 (%) (UNDP HDR) |
|---|---|---|---|---|---|---|---|---|
| People fully vaccinated per hundred, Feb. 2022 (OWID) | Pearson Correlation | 1 | 0.739** | 0.623** | 0.847** | N/A | | |
| | Sig. (2-tailed) | | 0.000 | 0.000 | 0.000 | | | |
| | N | 146 | 130 | 142 | 139 | | | |
| People fully vaccinated per hundred, July 2021 (OWID) | Pearson Correlation | 0.693** | 1 | 0.404** | 0.695** | N/A | | |
| | Sig. (2-tailed) | 0.000 | | 0.000 | 0.000 | | | |
| | N | 138 | 144 | 139 | 143 | | | |
| Human Development Index (HDI) 2019 | Pearson Correlation | 0.826** | 0.695** | 0.622** | 1 | N/A | | |
| | Sig. (2-tailed) | 0.000 | 0.000 | 0.000 | | | | |
| | N | 145 | 143 | 147 | 152 | | | |
| Life expectancy at birth 2019 (HDI component) | Pearson Correlation | 0.836** | 0.647** | 0.638** | 0.923** | N/A | | |
| | Sig. (2-tailed) | 0.000 | 0.000 | 0.000 | 0.000 | Component of HDI | | |
| | N | 146 | 144 | 148 | 152 | | | |
| Mean years of schooling 2019 (HDI component) | Pearson Correlation | 0.688** | 0.594** | 0.575** | 0.924** | N/A | | |
| | Sig. (2-tailed) | 0.000 | 0.000 | 0.000 | 0.000 | Component of HDI | | |
| | N | 145 | 142 | 147 | 150 | | | |
| Expected years of schooling 2019 (HDI component) | Pearson Correlation | .788** | .643** | 0.575** | .915** | N/A | | |
| | Sig. (2-tailed) | .000 | .000 | 0.000 | .000 | Component of HDI | | |
| | N | 142 | 139 | 147 | 147 | | | |
| GNI per capita 2019 in 2017 PPP (HDI component) | Pearson Correlation | 0.706** | 0.744** | 0.422** | 0.818** | N/A | | |
| | Sig. (2-tailed) | 0.000 | 0.000 | 0.000 | 0.000 | Component of HDI | | |
| | N | 142 | 139 | 147 | 147 | | | |
| Inequality in life expectancy 2015-2020 (HDR dataset) | Pearson Correlation | -0.793** | -0.641** | -0.668** | -0.936** | N/A | | |
| | Sig. (2-tailed) | 0.000 | 0.000 | 0.000 | 0.000 | Collinear with HDI | | |
| | N | 143 | 140 | 148 | 148 | | | |
| Inequality in education 2019 (HDR dataset) | Pearson Correlation | -0.663** | -0.530** | -0.584** | -0.847** | N/A | | |
| | Sig. (2-tailed) | 0.000 | 0.000 | 0.000 | 0.000 | Collinear with HDI | | |
| | N | 137 | 135 | 142 | 143 | | | |
| Inequality in income 2019 (HDR dataset) | Pearson Correlation | -0.286** | -0.337** | -0.302** | -0.378** | Partial Correlation | 0.037 | -0.099 |
| | Sig. (2-tailed) | 0.001 | 0.000 | 0.000 | 0.000 | Sig. (2-tailed) | 0.683 | 0.266 |
| | N | 126 | 123 | 130 | 131 | df | 123 | 127 |
| Gini Index 2019 (HDR dataset) | Pearson Correlation | -0.298** | -0.242** | -0.262** | -0.335** | Partial Correlation | -0.023 | -0.092 |
| | Sig. (2-tailed) | 0.001 | 0.006 | 0.002 | 0.000 | Sig. (2-tailed) | 0.799 | 293 |
| | N | 132 | 130 | 134 | 137 | df | 128 | 130 |
| PISA Score for Reading 2018 (OWID) | Pearson Correlation | 0.529** | 0.428** | 0.295* | 0.791** | Partial Correlation | 0.159 | 0.082 |
| | Sig. (2-tailed) | 0.000 | 0.000 | 0.016 | 0.000 | Sig. (2-tailed) | 0.209 | 0.518 |
| | N | 65 | 67 | 66 | 67 | df | 62 | 63 |
| PISA Score for Mathematics 2018 (OWID) | Pearson Correlation | 0.451** | 0.375** | 0.309* | 0.748** | Partial Correlation | 0.044 | 0.113 |
| | Sig. (2-tailed) | 0.000 | 0.002 | 0.011 | 0.000 | Sig. (2-tailed) | 0.727 | 0.367 |
| | N | 66 | 68 | 67 | 68 | df | 63 | 64 |
| PISA Score for Science 2018 (OWID) | Pearson Correlation | 0.513** | 0.379** | 0.285* | 0.732** | Partial Correlation | 0.168 | 0.080 |
| | Sig. (2-tailed) | 0.000 | 0.001 | 0.019 | 0.000 | Sig. (2-tailed) | 0.182 | 0.522 |
| | N | 66 | 68 | 67 | 68 | df | 63 | 64 |
| World Bank - Poverty ratio | Pearson Correlation | -0.641** | -0.580** | -0.560** | -0.670** | Partial Correlation | -0.225 | -0.261 |
| | Sig. (2-tailed) | 0.000 | 0.000 | 0.000 | 0.000 | Sig. (2-tailed) | 0.116 | 0.061 |
| | N | 51 | 51 | 50 | 51 | df | 48 | 47 |
| World Bank - National poverty ratio | Pearson Correlation | -0.637$^{**}$ | -0.494$^{**}$ | -0.536$^{**}$ | -0.705$^{**}$ | Partial Correlation | -0.167 | **-0.213$^{*}$** |
| | Sig. (2-tailed) | 0.000 | 0.000 | 0.000 | 0.000 | Sig. (2-tailed) | 0.066 | **0.018** |
| | N | 123 | 120 | 125 | 127 | df | 119 | **121** |
| Extreme poverty rate (OWID) | Pearson Correlation | -0.654 | -.468$^{**}$ | -0.389$^{**}$ | -.770$^{**}$ | Partial Correlation | -0.068 | 0.065 |
| | Sig. (2-tailed) | 0.000 | .000 | 0.000 | .000 | Sig. (2-tailed) | 0.496 | 0.521 |
| | N | 145 | 97 | 102 | 103 | df | 100 | 99 |
| Corruption Perception Index CPI 2020 (Transparency International) | Pearson Correlation | 0.689$^{**}$ | 0.663$^{**}$ | 0.480$^{**}$ | 0.766$^{**}$ | Partial Correlation | 0.135$^{*}$ | -0.004 |
| | Sig. (2-tailed) | 0.000 | 0.000 | 0.000 | 0.000 | Sig. (2-tailed) | 0.109 | 0.965 |
| | N | 144 | 142 | 146 | 150 | df | 140 | 142 |
| **Share of people who trust their national government 2018 (OWID, from Wellcome Global Monitor)** | **Pearson Correlation** | **0.053** | **-0.069** | 0.182$^{*}$ | **-0.066** | **Partial Correlation** | **0.217$^{*}$** | **0.266$^{*}$** |
| | Sig. (2-tailed) | 0.564 | 0.462 | 0.046 | 0.463 | Sig. (2-tailed) | 0.018 | 0.003 |
| | N | 120 | 116 | 121 | 124 | df | 117 | 118 |
| **Share of people who trust doctors and nurses in their country 2018 (OWID, from Wellcome Global Monitor)** | **Pearson Correlation** | 0.575$^{**}$ | **0.413$^{**}$** | 0.497$^{**}$ | **0.536$^{**}$** | **Partial Correlation** | **0.267$^{*}$** | **0.272$^{*}$** |
| | Sig. (2-tailed) | 0.000 | 0.000 | 0.000 | 0.000 | Sig. (2-tailed) | 0.002 | 0.002 |
| | N | 129 | 126 | 131 | 134 | df | 126 | 128 |
| Health expenditure % of GDP in 2017 (HDR dataset) | Pearson Correlation | 0.348$^{**}$ | 0.326$^{**}$ | 0.270$^{**}$ | 0.387$^{**}$ | Partial Correlation | 0.014 | 0.054 |
| | Sig. (2-tailed) | 0.000 | 0.000 | 0.001 | 0.000 | Sig. (2-tailed) | 0.868 | 0.523 |
| | N | 138 | 136 | 142 | 144 | df | 135 | 139 |
| Physicians per 1000 people 2019 (HDR dataset) | Pearson Correlation | 0.620$^{**}$ | 0.576$^{**}$ | 0.511$^{**}$ | 0.775$^{**}$ | Partial Correlation | -0.033 | 0.076 |
| | Sig. (2-tailed) | 0.000 | 0.000 | 0.000 | 0.000 | Sig. (2-tailed) | 0.698 | 0.368 |
| | N | 139 | 136 | 145 | 144 | df | 135 | 141 |
| Hospital beds per 1000 people 2019 (HDR dataset) | Pearson Correlation | 0.394$^{**}$ | 0.309$^{**}$ | 0.399$^{**}$ | 0.564$^{**}$ | Partial Correlation | -0.149 | 0.078 |
| | Sig. (2-tailed) | 0.000 | 0.000 | 0.000 | 0.000 | Sig. (2-tailed) | 0.092 | 0.375 |
| | N | 131 | 128 | 135 | 134 | df | 127 | 131 |
| **Cardiovascular death rate (OWID)** | Pearson Correlation | -0.497** | -.376** | -0.221** | -.410** | Partial Correlation | **-0.300** | 0.055 |
| | Sig. (2-tailed) | 0.000 | .000 | 0.008 | .000 | Sig. (2-tailed) | **0.000** | 0.520 |
| | N | 145 | 137 | 142 | 144 | df | **141** | 138 |
| Prevalence of diabetes (OWID) | Pearson Correlation | 0.238** | .120 | 0.125 | .269** | Partial Correlation | 0.031 | -0.056 |
| | Sig. (2-tailed) | 0.004 | .165 | 0.139 | .001 | Sig. (2-tailed) | 0.714 | 0.509 |
| | N | 144 | 136 | 141 | 144 | df | 141 | 138 |
| **Infants immunized for measles at 12 months, 2019 (%) (UNDP HDR)** | **Pearson Correlation** | **0.623**** | **0.404**** | 1 | **0.622**** | **Partial Correlation** | **0.231** | N/A |
| | Sig. (2-tailed) | 0.000 | 0.000 | | 0.000 | Sig. (2-tailed) | 0.006 | |
| | N | 142 | 139 | 148 | 147 | df | 138 | |

**. Correlation is significant at the 0.01 level (2-tailed).
*. Correlation is significant at the 0.05 level (2-tailed).

The three components of the HDI have differential predictive power for COVID-19 vaccination success (Table 3). A multiple regression model of the vaccination rate in February 2022 on the three dimensions of HDI (Model 1 includes the mean years of schooling, and Model 2 includes the expected years of schooling) indicates that, when controlling for the other dimensions, *the strongest predictor remains life expectancy*. The model, including all three HDI dimensions, does not lead to a substantial increase in predictive power. This is due to the fact that life expectancy, GNI per capita, and the mean and expected years of schooling are strongly intercorrelated and the latter do not contribute much in terms of additional explanatory power.

*The educational component of the HDI and GNI per capita are less powerful predictors than life expectancy in a multivariate model*. Either of education or GNI per capita may be statistically significant, but not both, depending on the chosen indicator for education (Model 1 and Model 2). The mean value of years of schooling in Model 1 is not a statistically significant predictor, but GNI per capita is. In Model 2 the expected value for years of schooling retains statistical significance, but GNI per capita does not. In Model 3, we see

that life expectancy is the strongest predictor for measles vaccination, followed by mean years of schooling. The same holds if we include expected years of schooling instead.

Table 3. Multiple regression model of the rate of people fully vaccinated in Feb. 2022 on HDI components: life expectancy, GNI per capita and mean years of schooling in 2019. Source: Authors' analysis of publicly available data from UNDP HDR and Our World in Data

| | **Model 1**<br>**Dependent variable: People fully vaccinated (%), February 2022** | | **Model 2**<br>**Dependent variable: People fully vaccinated (%), February 2022** | | **Model 3**<br>**Dependent variable: Infants immunized for measles at 12 months (%), 2019** | |
|---|---|---|---|---|---|---|
| **Independent variables:** | **Beta** | **Sig.** | **Beta** | **Sig.** | **Beta** | **Sig.** |
| Life expectancy at birth 2019 | **0.674**** | **0.000** | **0.522**** | **0.000** | **0.513** | **0.000** |
| GNI per capita 2019 (in 2017 PPP) | **0.206**** | **0.006** | 0.113 | 0.121 | -0.140 | 0.150 |
| Mean years of schooling 2019 | 0.003 | 0.968 | N/A | | **0.275** | **0.011** |
| Expected years of schooling | N/A | | **0.270**** | **0.002** | N/A | |
| Listwise N | | 142 | | 142 | | 147 |
| Adjusted R Square | | 0.700 | | 0.719 | | 0.408 |

Going back to partial correlations, *other indicators of educational outcome at country-level do not add predictive power beyond the HDI.* There are statistically significant bivariate correlations between vaccination rates and Programme for International Student Assessment (PISA) scores (Table 2). Still, the partial correlations for each of the PISA scores become statistically insignificant when controlling for HDI, life expectancy or GNI (PISA scores are only available for 67 countries). This indicates that, at country-level, literacies influence vaccination success insofar as they translate into higher life expectancy and GNI.

While a wide variety of indicators of human, social, and economic capital are correlated with vaccination success, both in July 2021 and February 2022, their predictive relevance is, most often, overlapping with the HDI. As we can see in Table 2, partial correlations when controlling for the HDI are, as a rule, statistically insignificant. Two indicators of social capital contribute to predicting vaccination success beyond the HDI: the share of people who trust doctors and nurses and the share of people who trust their national government. *Trust seems to play a significant role in the country-level success of the COVID-19 vaccination campaign.*

*Indicators of health system effectiveness retain statistically significant partial correlations with the vaccination rate in February 2022 when controlling for the HDI*. Cardiovascular (CVD) death rate has a partial correlation of –0.300 ($p$ = 0.000), and the proportion of infants immunized for measles before one year of age has a partial correlation of 0.231 ($p$ = 0.006). CVD are the leading cause of death globally. While their prevalence is higher in more developed countries, the associated mortality is higher in less developed countries. This makes this indicator a powerful proxy to capture the effectiveness of a country's medical system and overall social organization in increasing lifespan. The proportion of infants immunized for measles is a more specific indicator, pointing to a country's performance in its vaccination infrastructure. The prevalence of diabetes is not correlated with the COVID vaccination rate when controlling for the HDI, despite diabetes being a risk factor for severe COVID infections, which was associated with priority in the early vaccination campaigns.

*The pattern of correlations for predicting infant measles vaccination for one-year olds is very similar with COVID-19 vaccination.* The strongest bivariate predictors are life expectancy and the HDI. When controlling for the HDI, trust in the national government and trust in doctors and nurses remain statistically significant, but others indicators do not – except national poverty rates, which are relevant for measles but not for COVID-19 vaccination. Conversely, the CVD death rate remains significant for COVID-19 vaccination when controlling for the HDI, but not for measles.

In Table 4, we estimated a multiple linear regression of vaccination rates in February 2022 on HDI and the predictors that retained statistical significance when controlling for HDI: trust in doctors and nurses, trust in national government, infants immunized for measles, and the CVD death rate. In Model 3, the HDI remained the strongest predictor of vaccination success. The share of people who trust doctors and nurses and the trust in the national government are no longer statistically significant, when they are both included in the model. The other two health outputs remained statistically significant. We then excluded trust in the national government in Model 4, given that it correlates highly with trust in doctors and nurses. As a result of this model respecification, in Model 4, trust in doctors and nurses became statistically significant. In Model 4 the national poverty rate is also a marginally statistically significant predictor for COVID-19 vaccination.

*Table 4. Multiple regression model of vaccination rates on HDI, trust indicators, and cardiovascular death rate. Source: Authors' analysis of publicly available data from UNDP HDR and Our World in Data.*

| | **Model 3** | | **Model 4** | | **Model 5** | |
|---|---|---|---|---|---|---|
| | **Dependent variable: Covid-19 vaccination rate in February 2022** | | | | **Dependent variable: Infants immunized for measles at 12 months, 2019 (%)** | |
| | **Standardized coefficient Beta** | **Sig.** | **Standardized coefficient Beta** | **Sig.** | **Standardized coefficient Beta** | **Sig.** |
| **Independent variables:** | | | | | | |
| HDI 2019 (HDR dataset) | 0.588$^{**}$ | 0.000 | **0.472$^{**}$** | **0.003** | **0.389$^{**}$** | **0.001** |
| World Bank National poverty rate | -0.030 | 0.677 | **-0.145$^{*}$** | **0.044** | -0.109 | 0.283 |
| Share of people who trust doctors and nurses in their country, 2018 (OWID) | 0.177 | 0.065 | **0.150$^{*}$** | **0.021** | **0.287$^{**}$** | **0.001** |
| Cardiovascular death rate | -0.197$^{**}$ | 0.000 | **-0.219$^{**}$** | **0.000** | 0.098 | 0.216 |
| **Infants immunized for measles at 12 months, 2019 (%) (UNDP HDR)** | 0.135$^{**}$ | 0.020 | **0.143$^{**}$** | **0.038** | Dependent variable | |
| Share of people who trust their national government, 2018 (OWID) | -0.001 | 0.094 | Not included | | Not included | |
| Adjusted R Square | 0.753 | | 0.703 | | 0.395 | |
| Listwise N | 105 | | 111 | | 111 | |

As discussed before, a similar understanding holds for measles vaccination (Table 4, Model 5). The HDI is also the strongest predictor of the rate of infants immunized for measles. The lower beta coefficient also reflects the nonlinear relationship, which is better approximated by a logarithmic curve, because of the vaccination plateau (Fig. 3). Therefore, the predictive relevance of the HDI goes beyond COVID-19 vaccination, covering previous, better institutionalized vaccines as well. The rate of trust in doctors and nurses is also a significant predictor of measles vaccination. The CVD rate does not add a statistically significant predictive power for measles vaccination. Neither does the national poverty rate, despite having a significant partial correlation when controlling for the HDI.

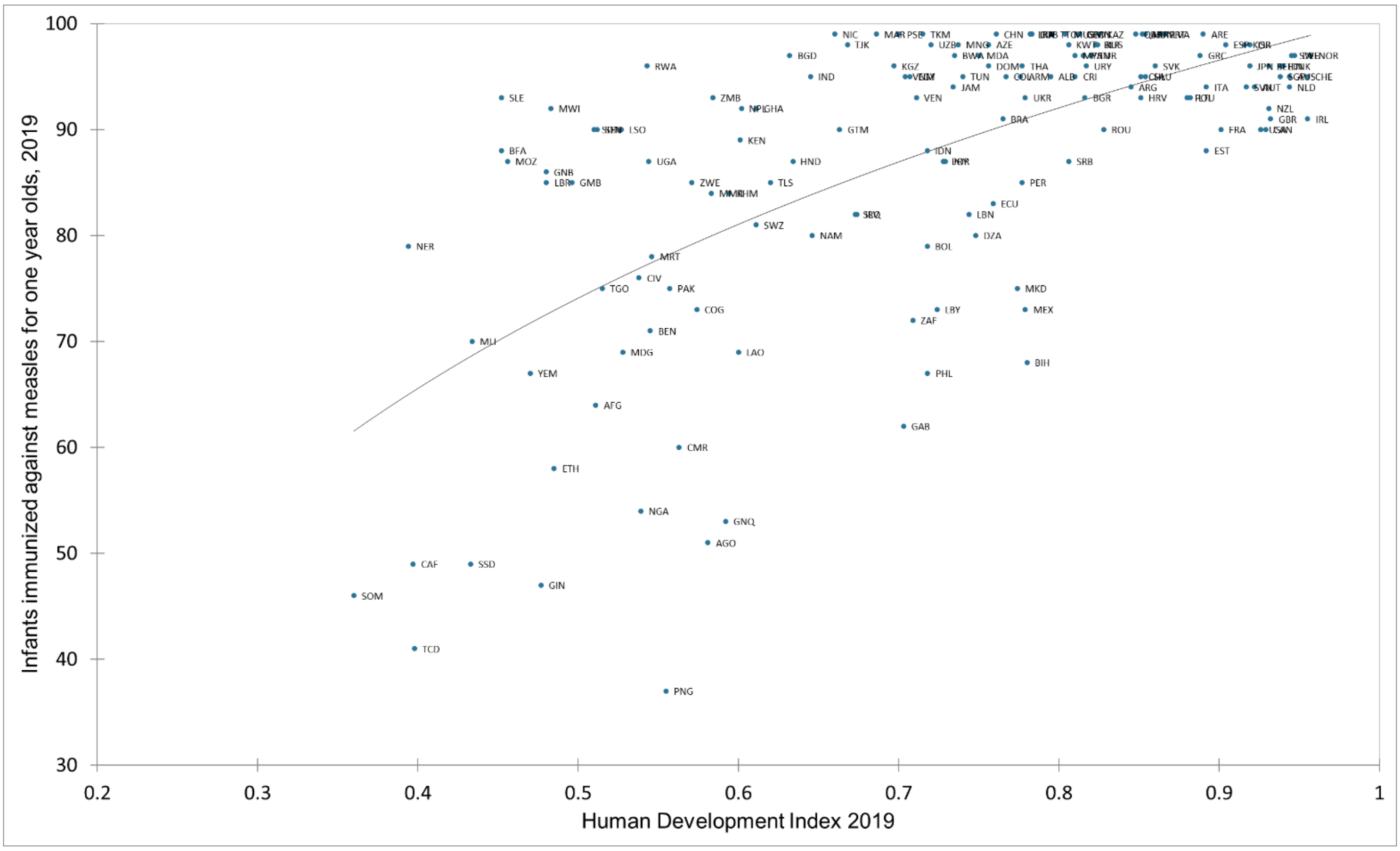

**Fig. 3.** Scatterplot of rates of infants immunized against measles for one-year olds, 2019, vs. HDI 2019, with a logarithmic growth trendline. Source: Authors' analysis on data from UNDP Human Development Reports.

The relationship between COVID-19 vaccination rates and trust in doctors and nurses, while controlling for HDI and other country-level health outcomes, is useful to clarify divergences that rank prominently in public debate. COVID-19 vaccination trajectories among countries in the same HDI categories have been quite different. While the United States and Israel were initially champions due to securing early access, by July 2021 they had started losing ground compared with other high HDI countries, which benefit from very high levels of trust in their medical systems [33], [34] (Fig. 4).

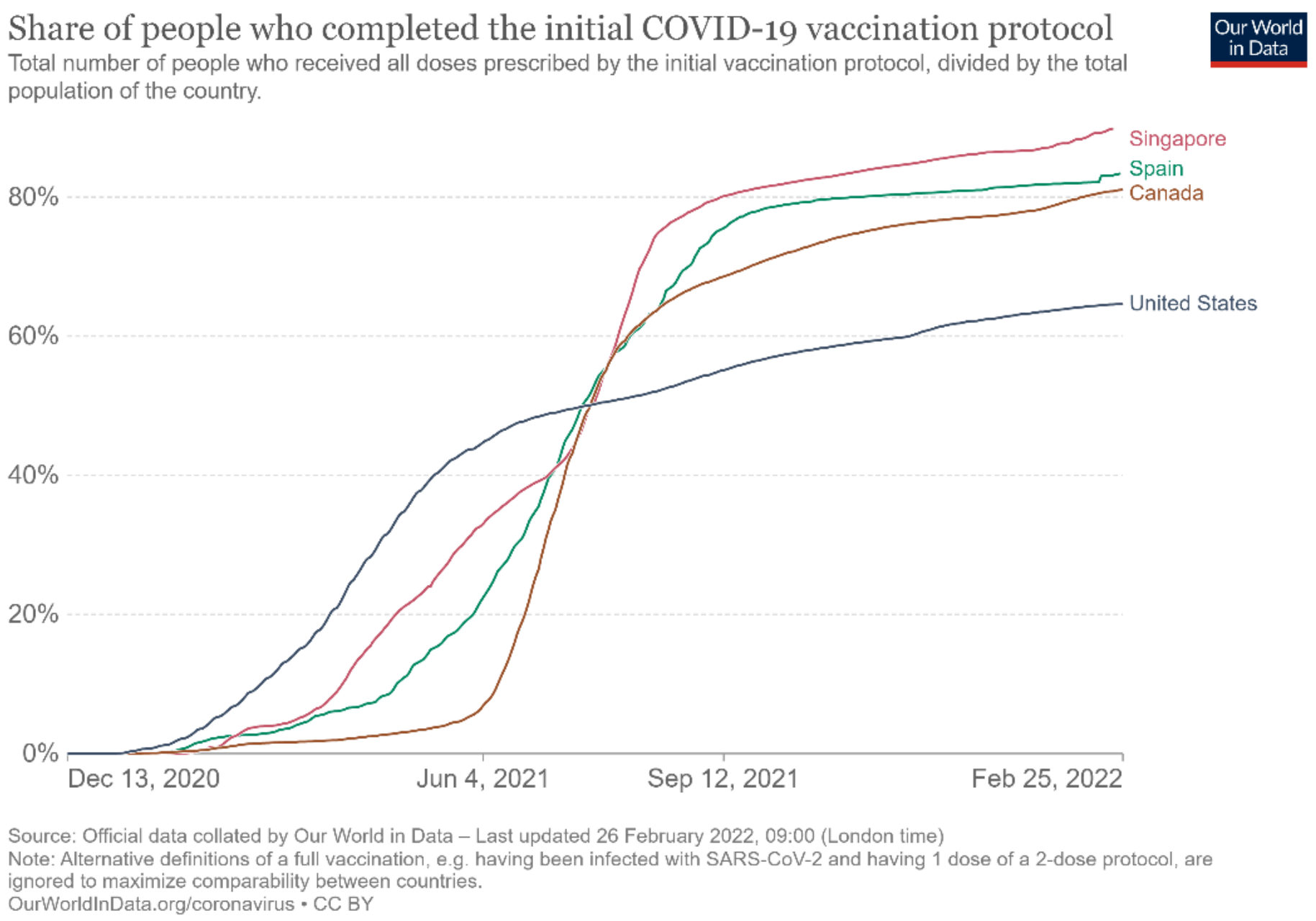

**Fig. 4.** Rate of fully vaccinated people in Singapore, Spain, Canada, and the US. Source: Our World in Data.

Another contrasting story concerns Romania and Spain. Romania has had a temporary advantage in the early days of the vaccination campaign, which was lost in June, when Spain started gaining ground, reaching one of the top global vaccination rates in February 2022. Since both are members of the European Union, this difference reflects vaccine hesitancy more than vaccine inequity. Trust in the healthcare system has been invoked to account for differences in vaccine hesitancy. Spain is credited with a high level of trust in vaccines and in its medical system [34], [33]. On the other hand, less than 40% of the Romanian public trusts public hospitals [35], following a decade-long struggle with corruption [36]. India, which has a very high rate of trust in doctors and nurses [33], has reached higher vaccination rates than Romania, despite its considerable challenges in vaccine availability. The low levels of public trust in the medical system, associated with a longstanding crisis [37], [34], also seem to account, in part, for Bulgaria's low rate of vaccination despite the high availability of vaccines typical of EU countries (Fig. 5).

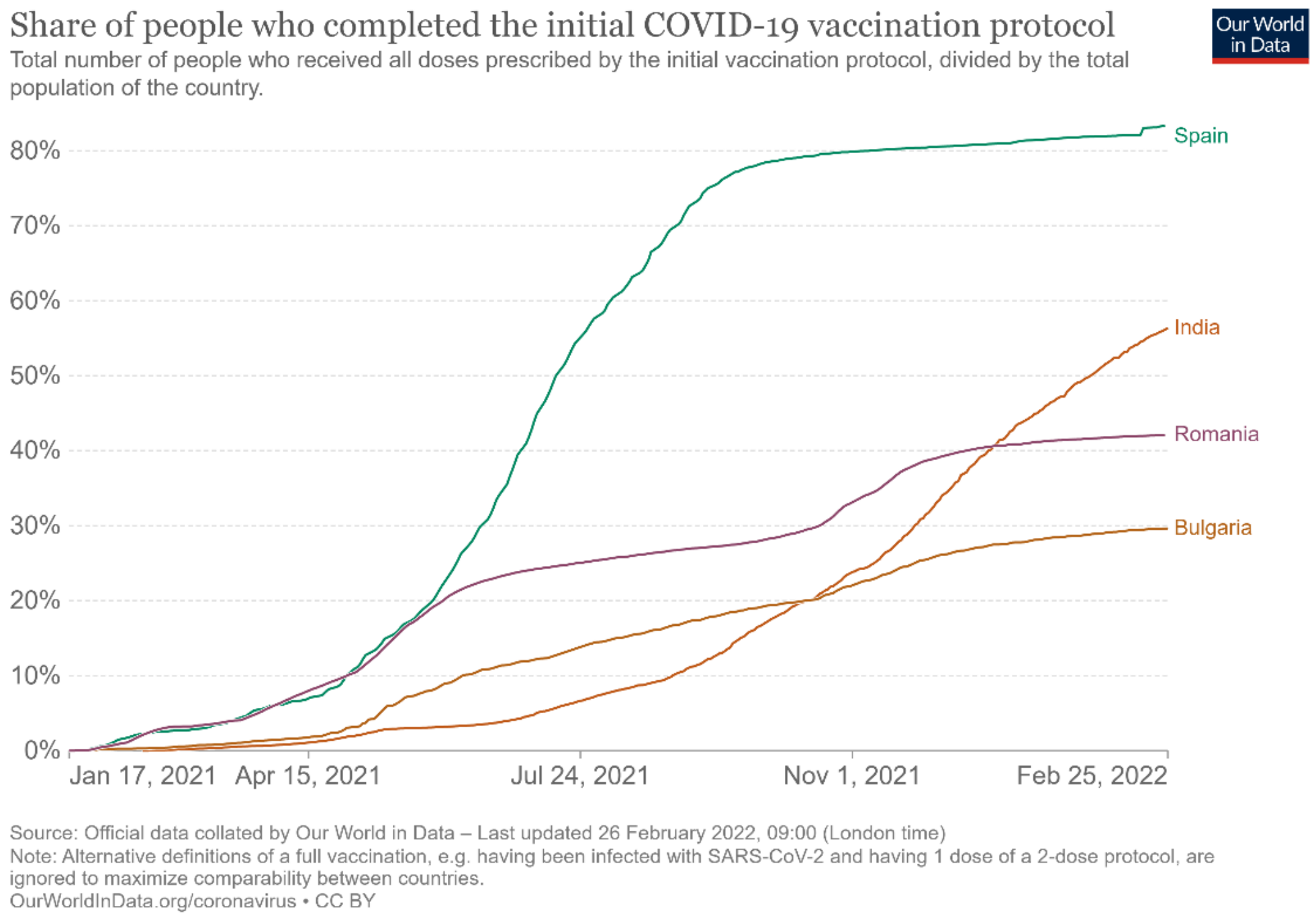

**Fig. 5.** Rate of fully vaccinated people in Spain, India, Romania, and Bulgaria. Source: Our World in Data.

# Discussion

Our exploratory analysis of social patterns that uphold vaccination success, in the case of COVID-19 and measles, highlights *the role of the HDI as the strongest predictor* among a set of widely used measures of human, social, and economic capital. Among its three dimensions, *life expectancy is the most relevant for accounting for COVID-19 and measles vaccination success*. It is important to keep in mind that these are synthetic indicators of the effectiveness of social organizations, reflecting the country-level outcomes of the interplay of local and global forces of creation and destruction.

*Education outcomes, measured through mean years of schooling, expected years of schooling, or PISA results, add less explanatory power than life expectancy*, in regard to the COVID-19 and measles vaccination. This supports the argument that vaccination success is less a matter of overcoming deficits in scientific literacy, and more a matter of establishing public trust in a health system and science with proven anterior performance in keeping people healthy and alive [10].

*Our analysis also highlights the role of trust in doctors and nurses as a predictor of vaccination success*. It remains statistically significant when controlling for the HDI and other generic and specific indicators of health system effectiveness (CVD mortality and measles vaccination coverage, respectively). Trust is statistically significant in partial correlation and multiple regression models of both COVID-19 and measles vaccination, while other indicators concerning economic inequality, perceived corruption, and inputs into the health system do not add predictive value beyond the HDI. National poverty rates seem to remain a relevant predictor for both types of vaccination, though statistical significance is oscillating around the 5% threshold, depending on model specification.

COVID-19 vaccines prove to be part of the Matthew effect of accumulating advantages and exacerbating disadvantages that the pandemic inflicted on societies and communities across the world [38]. At the same time, the remaining 28% of variability that cannot be determined through these sets of metrics points to a considerable scope for collective and individual agency in a time of crisis. For example, countries with an HDI of approximatively 85 ranged from rates of 40% to 80% for fully vaccinated people. The mobilization and coordination in the vaccination campaigns of citizens, medical professionals, scientists, journalists, and politicians, among others, account for at least some of this variability in overcoming vaccine hesitancy and inequity.

# Author contribution

All authors made a significant contribution to the development of this manuscript and approved the final version for submission.

# Funding

This work was supported by the Ministry of Innovation, Scientific Research, and Digitalization, Romania, project PN-III-P4-ID-PCE-2020-1589.

## Declaration of Competing Interest

The authors declare that they have no known competing financial interests or personal relationships that could have appeared to influence the work reported in this paper.